\documentclass[prl,aps,amssymb,showpacs,twocolumn]{revtex4-1}

\usepackage{amsmath}
\usepackage{amssymb}
\usepackage{amsthm}
\usepackage{amsfonts}
\usepackage{listings}
\usepackage{enumerate}
\usepackage{latexsym}
\usepackage{graphicx}
\usepackage{psfrag}
\usepackage{psfrag}
\usepackage{bm}

\usepackage{graphicx}
\usepackage{subfigure}
\usepackage{graphics}
\usepackage{color}

\let\oldmarginpar\marginpar
\renewcommand{\marginpar}[1]{\-\oldmarginpar[\color{red}\raggedleft\tiny #1] {\color{red}\raggedright\tiny #1}}

\newcommand{\beq}{\begin{equation}}
\newcommand{\eneq}{\end{equation}}
\input{epsf}

\begin{document}

\tolerance 10000

\newcommand{\vk}{{\bf k}}

\title{Haldane Statistics in the Finite Size Entanglement Spectra of $1/m$ FQH States}

\author{M. Hermanns$^1$, A. Chandran$^1$, N. Regnault$^2$, B. Andrei Bernevig$^1$} 
\affiliation{$^1$ Department of Physics,
  Princeton University, Princeton, NJ 08544} 
\affiliation{$^2$ Laboratoire Pierre Aigrain, ENS and CNRS, 24 rue Lhomond, 75005 Paris, France}

\begin{abstract}
We conjecture that the counting of the levels in the orbital entanglement spectra (OES) of finite-sized Laughlin Fractional Quantum Hall (FQH) droplets at filling $\nu=1/m$ is described by the Haldane statistics of particles in a box of finite size. This principle explains the observed deviations of the OES counting from the edge-mode conformal field theory counting and directly provides us with a topological number of the FQH states inaccessible in the thermodynamic limit--- the boson compactification radius. It also suggests that the entanglement gap in the Coulomb spectrum in the conformal limit protects a universal quantity--- the \emph{statistics} of the state. 
We support our conjecture with ample numerical checks.
\end{abstract}
\date{\today}


\maketitle

Topological phases of matter have received renewed attention in the past several years due to advances in experimental techniques such as quasiparticle interferometry in FQH states \cite{Ainterf,NAinterf}. Unfortunately, determining if the ground state of a realistic Hamiltonian describing an experimentally observable system is topologically ordered \cite{toporder} remains an unsolved problem. The difficulties are multiple: be it identifying a set of \emph{universal} features (such as adiabatic transport quantities) that uniquely define a topological state, extrapolating finite-size calculations to the thermodynamic limit, or  extracting sub-leading terms (such as  topological entanglement entropy\cite{levinwen, kitaevpreskill}).

Entanglement between different parts of a system has emerged as a leading diagnostic of the topological nature of a many-body state, see for instance Refs. \cite{calabrese-04jsmp06002,haque2007, Dong-jh,PhysRevLett.104.130502,turner2010prb241102,EEphasetransitions}. Li and Haldane \cite{LiHaldane} discovered that the number of the first few eigenvalues in the orbital entanglement spectrum (OES) of model FQH states at filling $\nu$ is the counting of the conformal field theory associated with the edge-mode of the state in the thermodynamic limit. The same counting occurs in the low-lying OES of the exact ground state of the Hamiltonian with Coulomb interaction at filling $\nu$, thus justifying its interpretation as the `topological imprint' of the state.

However, the counting of the spectrum of a model FQH state quickly develops finite-size effects which are thought to have no structure. These finite-size levels strongly mix with the spurious levels higher in the OES of the Coulomb ground state and interfere with the determination of a low-lying universal spectrum. Using a flat-band procedure called the conformal limit, the non-universal part of the Coulomb spectrum can be completely separated from a low-lying part with the \emph{same} counting as the finite-size OES of the model FQH state at the same filling by a \emph{full} entanglement gap\cite{entgap}. This suggests a counting principle behind the finite-size level counting of the model states.

In the present paper, we conjecture a counting principle for the finite-size spectra of the Laughlin $\nu=1/m$ states. When the system is cut in orbital space, the number of non-zero eigenvalues of the reduced density matrix as a function of the angular momentum of subsystem $A$ exhibits Haldane exclusion statistics \cite{haldanestats} of a boson of compactification radius $\sqrt{m}$ quantized in a box of finite orbital length. The conjecture predicts the observed counting of the \emph{full} entanglement spectrum of the $m=2,3$ states and most of the counting of the spectra of the $m>3$ states. The existence of such a counting principle lends meaning to the OES at finite size and suggests a new interpretation of the entanglement gap in the Coulomb spectrum, known to be non-zero in the thermodynamic limit from numerical studies \cite{entgap}, as protecting the Haldane statistics of the phase. 

 Our counting principle also shows that the finite-size OES resolves more than just the central charge of the edge theory: it provides us with a new and simple way of extracting the boson compactification radius, previously determined by intricate scaling arguments \cite{pasquier, calabresecardy2009}. 
In the thermodynamic limit the counting of modes of a U(1) boson does not depend on the compactification radius and only allows for determining the central charge of the CFT describing the edge physics. For example, the number of excitations of the bosonic and fermionic system are the same in the thermodynamic limit. However, a finite-size box probes the non-trivial exclusion statistics of the particles encoded in the parameter $m$. 
The finite-size OES thus determines \emph{all} the quantum numbers of the $1/m$ Laughlin states, and thereby---  as long as the entanglement gap is finite--- all the topological properties of the Coulomb state at the same filling.

The results that we present in this article hold on any surface of genus 0 pierced by $N_\phi$ flux quanta; for simplicity, let us choose the sphere geometry. The single particle states of each Landau level are eigenstates of $\hat{L}_z$, the $z$ component of angular momentum, with values ranging from $N_\phi/2$ (North pole) to $-N_\phi/2$ (South pole) in the lowest Landau level. Fermionic (bosonic) many-body wave functions of $N$ particles and total angular momentum $L_z^{tot}$ can be expressed as linear combinations of Fock states in the occupancy basis of the single particle orbitals. Each Fock state can be labeled either by $\lambda$, a partition of $L_z^{tot}$, or the occupation number configuration $n(\lambda)=\{n_j(\lambda), j=N_\phi/2,\ldots -N_\phi/2\}$, where $n_j(\lambda)$ is the occupation number of the single particle orbital with angular momentum $j$. The coefficient of every partition in the expansion of the $1/m$ Laughlin wave function can be obtained from that of a single `\emph{root}' partition $\lambda_0$ \cite{bernevighaldane2007} with $L_z^{tot}=0$ and occupation number configuration $n(\lambda_0)=\{10^{m-1} 10^{m-1} 1 \ldots\}$. $0^{m-1}$ denotes $m-1$ consecutive, unoccupied orbitals. $n(\lambda_0)$ is  $(1,m)$-admissible, i.e. it contains no more than one particle in $m$ consecutive orbitals. 
There is a one-to-one correspondence between the $(1,m)$-admissible configurations of $N$ particles in $(N_\phi+1)$ orbitals and the number of Laughlin (quasihole) states at filling $\nu=1/m$ and total flux $N_\phi$ via Jack polynomials \cite{bernevighaldane2007}. 

To obtain the orbital entanglement spectrum (OES), we cut the sphere between two adjacent orbitals after $l_A$ orbitals from the North pole (part $A$) or  $l_B=N_\phi +1-l_A$  orbitals from the South pole (part $B$). Let part $A$ be the smaller sub-system with $l_A\leq l_B$. Any many-body wave function can be expanded as $|\psi\rangle = \sum_{i,j} C_{ij} |\psi_A^i\rangle \otimes |\psi_B^j\rangle$,
where $|\psi_A^i\rangle $ ($|\psi_B^j\rangle$) is a basis of the Hilbert space of part $A$ ($B$).  
$C=(C_{ij})$ is the orbital entanglement matrix (OEM) and $\rho_A=C C^T $ is the reduced density matrix of $A$. Both are block-diagonal in the number of particles, $ N_A$ ($N_B$), and the total angular momentum, $L_z^A$ ($L_z^B$), of part $A$ ($B$). 
The OES is the plot of $\xi$, the negative logarithm of the eigenvalues of the sub-matrix of $\rho_A$ at fixed $N_A$, as a function of $L_z^A$. 
The OES level counting is determined by either $C$ or $\rho_A$, as they have identical rank.

We define $\Delta L_z = L_{z,max}^A - L_z^A$, where $L_{z,max}^A = m N_A(N-N_A)/2$ is the $z$-angular momentum of the configuration where the particles in $A$ are maximally close to the North pole. In the thermodynamic limit ($l_A \rightarrow \infty$ before $N_A\rightarrow\infty$), the number of levels in the OES for any $m$ grows as $(1,1,2,3,5,7,11\ldots )$ for $\Delta L_z=0,1,2\ldots$, matching the number of excitations of a chiral $U(1)$ boson at each $\Delta L_z$. Corrections to this counting occur because of the finite number of particles or orbitals in $A$. The OES counting for a finite $N_A$ as $l_A \rightarrow \infty$ is different from the above $U(1)$ counting, but remains the same for all $1/m$ Laughlin states. Dependence of the OES counting on $m$ arises only when  $l_A$ is also finite. 
\begin{figure}[tbp]
\centering
\includegraphics[width=6cm]{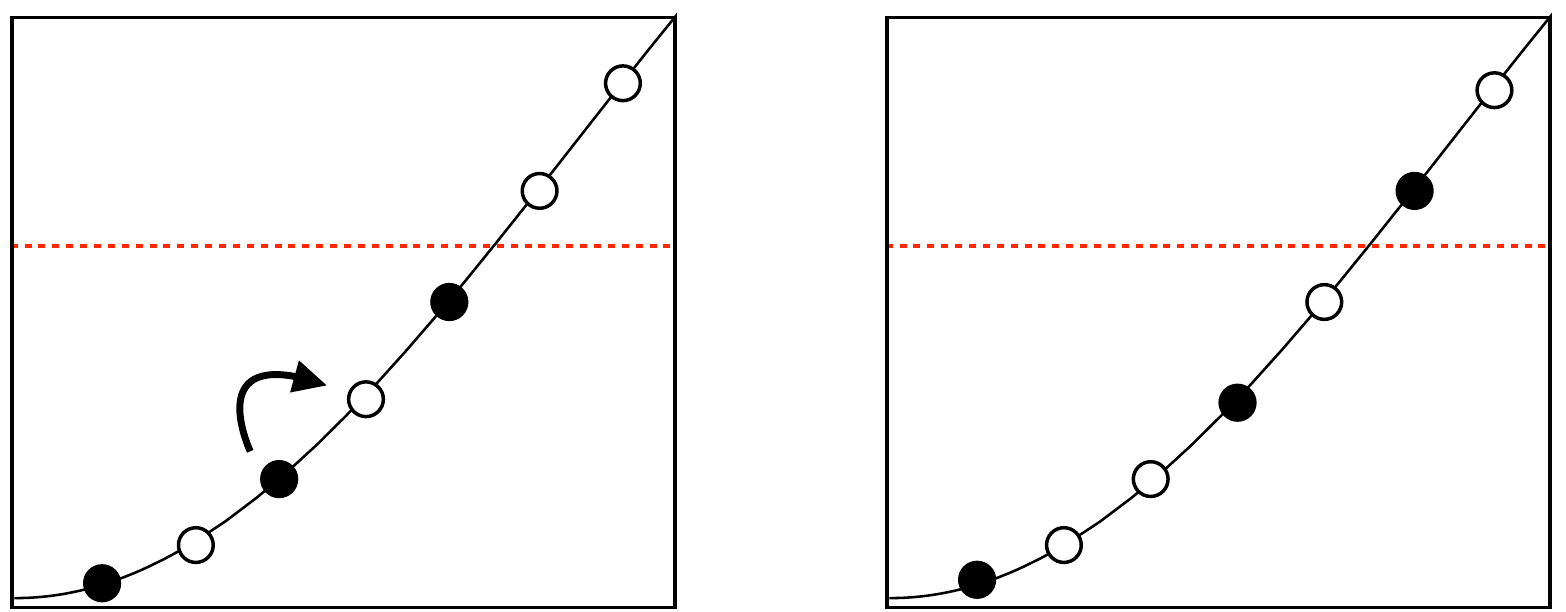}
\caption{Schematic indicating the excitations above the $\nu=1/2$ ground state that count the number of levels in the OES with $N_A=3$ in $l_{enl}=7$ orbitals. Haldane statistics implies that each occupied orbital `blocks' its two neighboring orbitals. Thus, only the excited state in the right box contributes to the level counting of the OES at $\Delta L_z=1$, while the configuration arising out of the transition (arrow) in the left box does not. }
\label{fig:bcomrad}
\end{figure}

Let us assume for the moment that $B$ is infinite, i.e. $l_B,N_B\rightarrow\infty$. We argue that the level counting of the OES exhibits Haldane exclusion statistics by identifying all the configurations in $A$ with a set of $(1,m)$-admissible configurations with the same quantum numbers. 
The latter obey the generalized Pauli principle that there is no more than one particle in $m$ consecutive orbitals corresponding to the statistical interaction $g=m$\cite{haldanestats}. 
Observe that the orbital cut at $l_A$ imposes a `hard-wall' potential on the subsystem $A$; it forbids the occupation of any orbital with $L_z\geq l_A$. 
For values of $l_A$ and $\Delta L_z$ such that none of the configurations in $A$ probe this wall, the $(1,m)$-admissible configurations count the number of levels in the OES \cite{Chandran2011}. 
For all other values, we conjecture that the $(1,m)$-admissible configurations of $N_A$ particles continue to count the levels in the OES, if we move the hard wall and increase the orbital size of $A$ to $l_{enl}>l_A$. 
A schematic view of this counting principle is shown in Fig.~\ref{fig:bcomrad}.
To determine $l_{enl} $, we use the fact that there is exactly one level in the OES at the minimal possible angular momentum, $L_{z,min}^A$. 
This is because there is only one allowed configuration in $A$ - $\{0\ldots 0 N_A \, | \, \ldots \}$ for bosonic Laughlin states and $\{0\ldots 01\ldots 1\, |\, \ldots \}$ for the fermionic ones, where `$|$' denotes the orbital cut.  
We therefore fix $l_{enl}$ such that there is \emph{exactly one} $(1,m)$-admissible partition at angular momentum $L_{z,min}^A$ and none at smaller angular momenta: $l_{enl}=l_A+m(N_A-1)/2$ (bosons) and $l_{enl}=l_A+(m-1)(N_A-1)/2$ (fermions).

%\paragraph{Conjecture}
In a numerically accessible system, both $A$ and $B$ are of finite size. The arguments in the previous paragraph then apply to both subsystems. We thus conjecture an upper bound $\mathcal{N}_{s}(l_A, N_A, \Delta L_z)$ to the level counting of the OES at given $N_A$, $l_A$, and $\Delta L_z$, taking both subsystems into account:
\begin{multline}
\label{eq:countform}
\mathcal{N}_{s}(l_A, N_A, \Delta L_z)\\= \textrm{min}[\mathcal{N}(l_A, N_A, \Delta L_z),\mathcal{N}(l_B, N_B, \Delta L_z)] ,
\end{multline} 
where $\mathcal{N}(l_a, N_a, \Delta L_z)$ for $a=A,B$ is the number of $(1,m)$-admissible states of $N_a$ particles in $l_{enl,a}=l_a+m(N_a-1)/2$ (bosons) or $l_{enl,a}=l_a+(m-1)(N_a-1)/2$ (fermions) orbitals at angular momentum $\Delta L_z$. It is defined by the generating function:
\begin{align}
\label{eq:genfunc}
\sum_{\Delta L_z=0}^{N_a \cdot N_a^h} {\cal N}(l_a, N_a, \Delta L_z) q^{\Delta L_z} &= \frac{(q)_{N_a+N_{a}^h}} {(q)_{N_a}(q)_{N_{a}^h}}\, ,
\end{align}
with $(q)_n=\prod_{i=1}^n (1-q^i)$ and  $N_a^h=l_{enl,a}-m(N_a-1)-1$. ${\cal N}(l_a, N_a, \Delta L_z)$ is a well-known quantity; it is the number of linearly independent Laughlin states of $N_a$ particles with $N_a^h$ added flux quanta at angular momentum $\Delta L_z$ \cite{RR96}. 
For nearly all values of $N_A,l_A$, and $\Delta L_z$, the bound is saturated and the observed counting is given by $\mathcal{N}_{s}(l_A, N_A, \Delta L_z)$. Specifically, Eq.~\eqref{eq:countform} predicts the correct level counting for the \emph{entire} OES for $m=2,3$, and for \emph{most} $(N_A, l_A)$ sectors of the $m>3$ states. 

%\paragraph{counting exact}
For given $(N_A,l_A)$, Eq.~\eqref{eq:countform} predicts the observed counting of the full spectrum if $\mathcal{N}_{s}(l_A, N_A, \Delta L_z)$ simplifies to $\mathcal{N}(l_A, N_A, \Delta L_z)$ \emph{for all} $\Delta L_z$ (or alternatively if it simplifies to  $\mathcal{N}(l_B, N_B, \Delta L_z)$ for all $\Delta L_z$). In other words, Eq.~\eqref{eq:countform} holds exactly when the corrections to the thermodynamic counting are only due to the finite-size of a single subsystem. 
For the bosonic states, a necessary and sufficient condition for this is (\textbf{i}) $N_B\geq \textrm{min}(N_A, N_A^h)$. 
For fermionic states, an additional condition (\textbf{ii}) $l_A-N_A\leq (m-1)(N-1)/2$ has to be imposed as a consequence of the Pauli exclusion statistics of the fermions in part $B$. For $m=2,3$, restricting the orbital cut to $l_A\leq N_\phi/2$ ensures that these conditions are satisfied. For $m>3$, both conditions can always be satisfied by choosing the system size to be sufficiently large. 

%\paragraph{upper bound}
When $N_B< \textrm{min}(N_A, N_A^h)$ ((\textbf{i}) does not hold), $\mathcal{N}_{s}(l_A, N_A, \Delta L_z)$ simplifies to $\mathcal{N}(l_B, N_B, \Delta L_z)$ at small $\Delta L_z$ and to $\mathcal{N}(l_A, N_A, \Delta L_z)$ at large $\Delta L_z$. 
If (\textbf{ii}) is not satisfied, the situation is reversed--- the level counting at small $\Delta L_z$ is equal to $\mathcal{N}(l_A, N_A, \Delta L_z)$, while the number of levels at large $\Delta L_z$ is equal to $\mathcal{N}(l_B, N_B, \Delta L_z)$. 
At both ends of the spectrum, the level counting of the OES is equal to $\mathcal{N}_{s}(l_A, N_A, \Delta L_z)$. 
In fact, Eq.~\eqref{eq:countform} is the observed level counting everywhere except possibly at a few values around $\Delta L_z^0$, when $\mathcal{N}(l_A,N_A,\Delta L_z^0)\approx \mathcal{N}(l_b,N_B,\Delta L_z^0)$. 
Near $\Delta L_z^0$, the counting is dependent on the finite sizes of $A$ \emph{and} $B$ and Eq.~\eqref{eq:countform} is only an upper bound.
We have tested our conjecture for all possible orbital cuts at all numerically accessible system sizes, i.e up to $N=(16,15,11,11,9,9,7,7)$ particles for $m=(2,3,\ldots, 9)$.

%\paragraph{Example}
\begin{figure*}[t]
%Laughlin 1/2 na<n_natcut
  \begin{minipage}[l]{0.245\linewidth}
   \includegraphics[width=\linewidth]{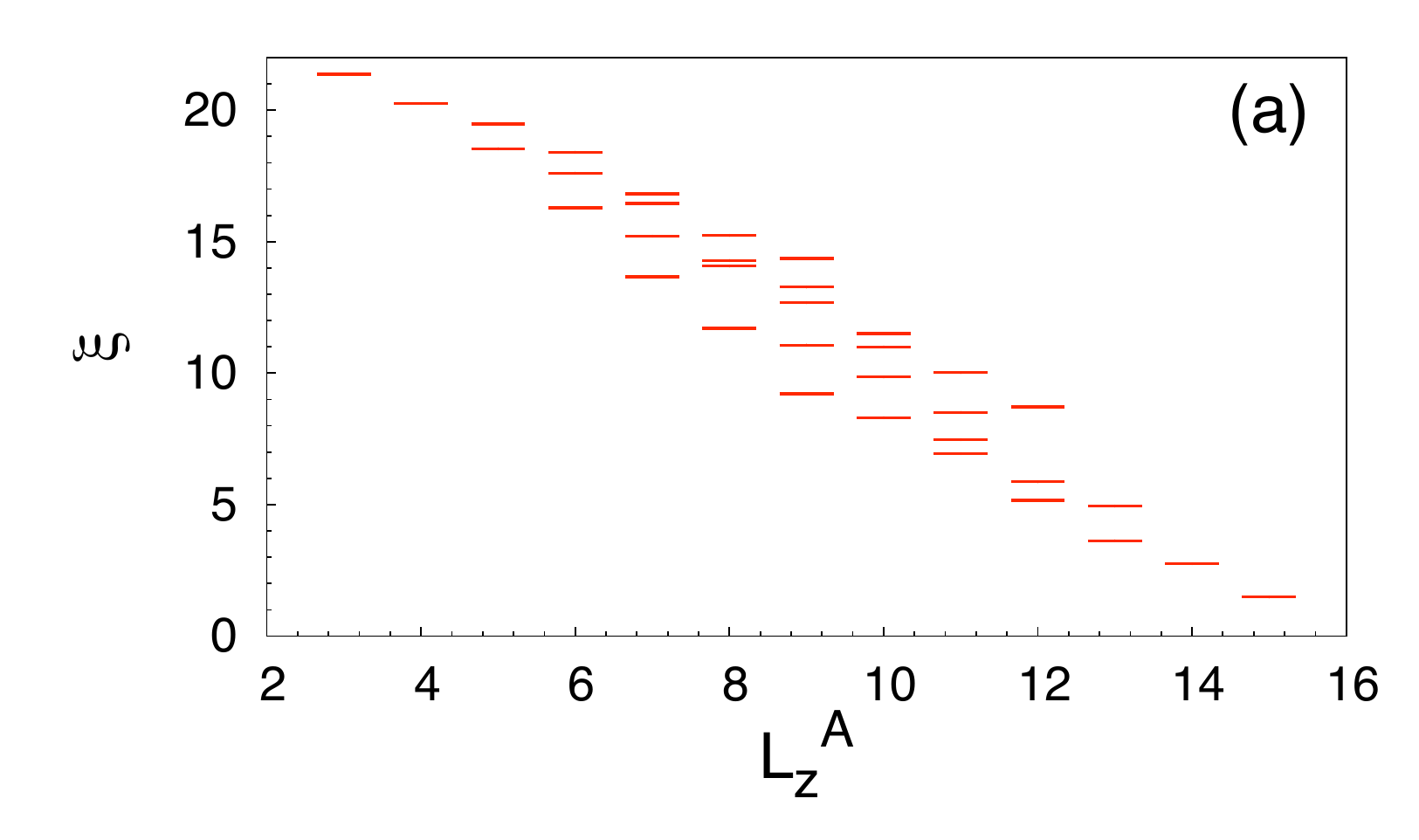}
 \end{minipage}
%\hspace{5pt}
%Laughlin 1/2 na=n natcut
 \begin{minipage}[l]{0.245\linewidth}
   \includegraphics[width=\linewidth]{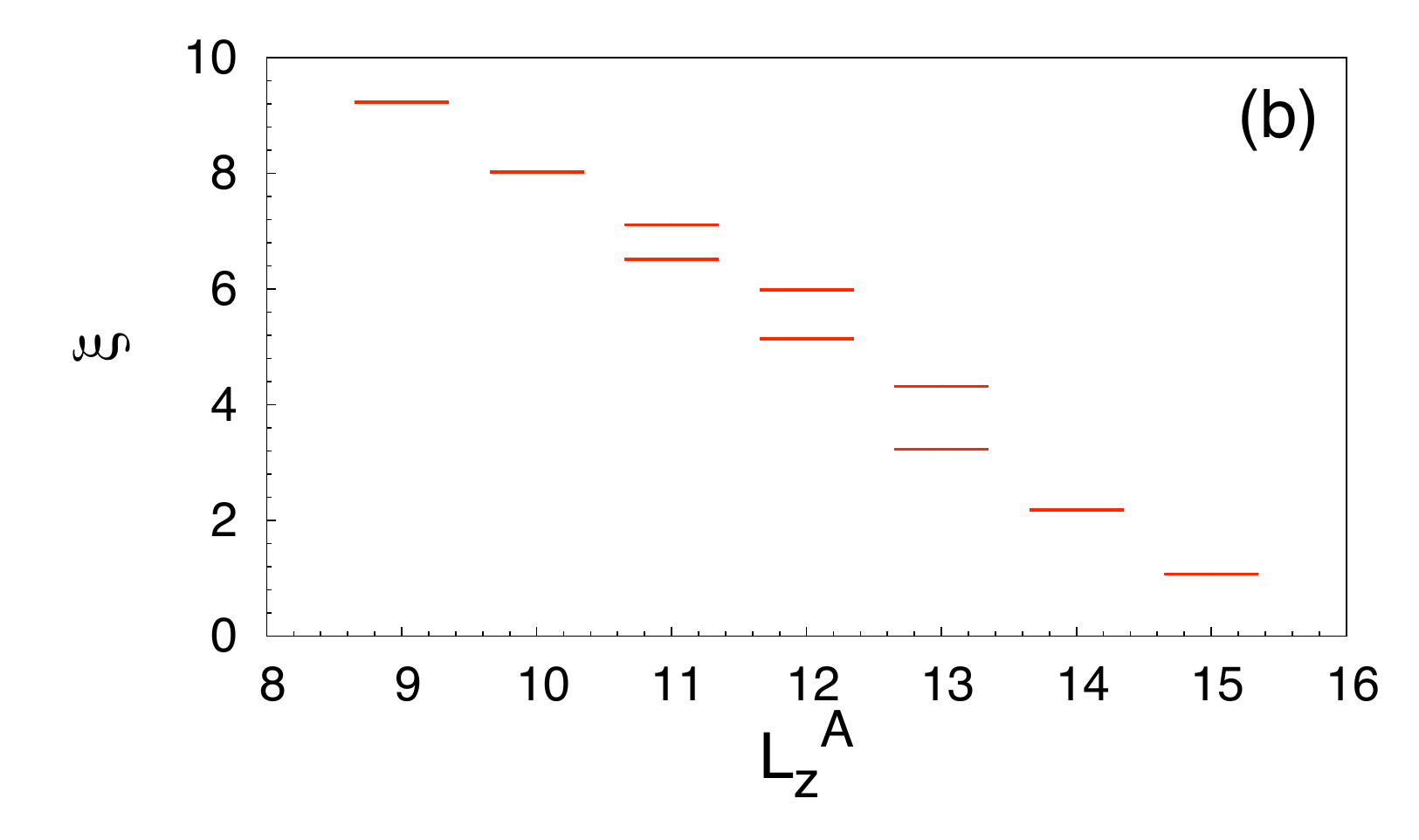}
 \end{minipage} 
%\hspace{5pt}
%Laughlin 1/2 na> n natcut
\begin{minipage}[l]{0.245\linewidth}
   \includegraphics[width=\linewidth]{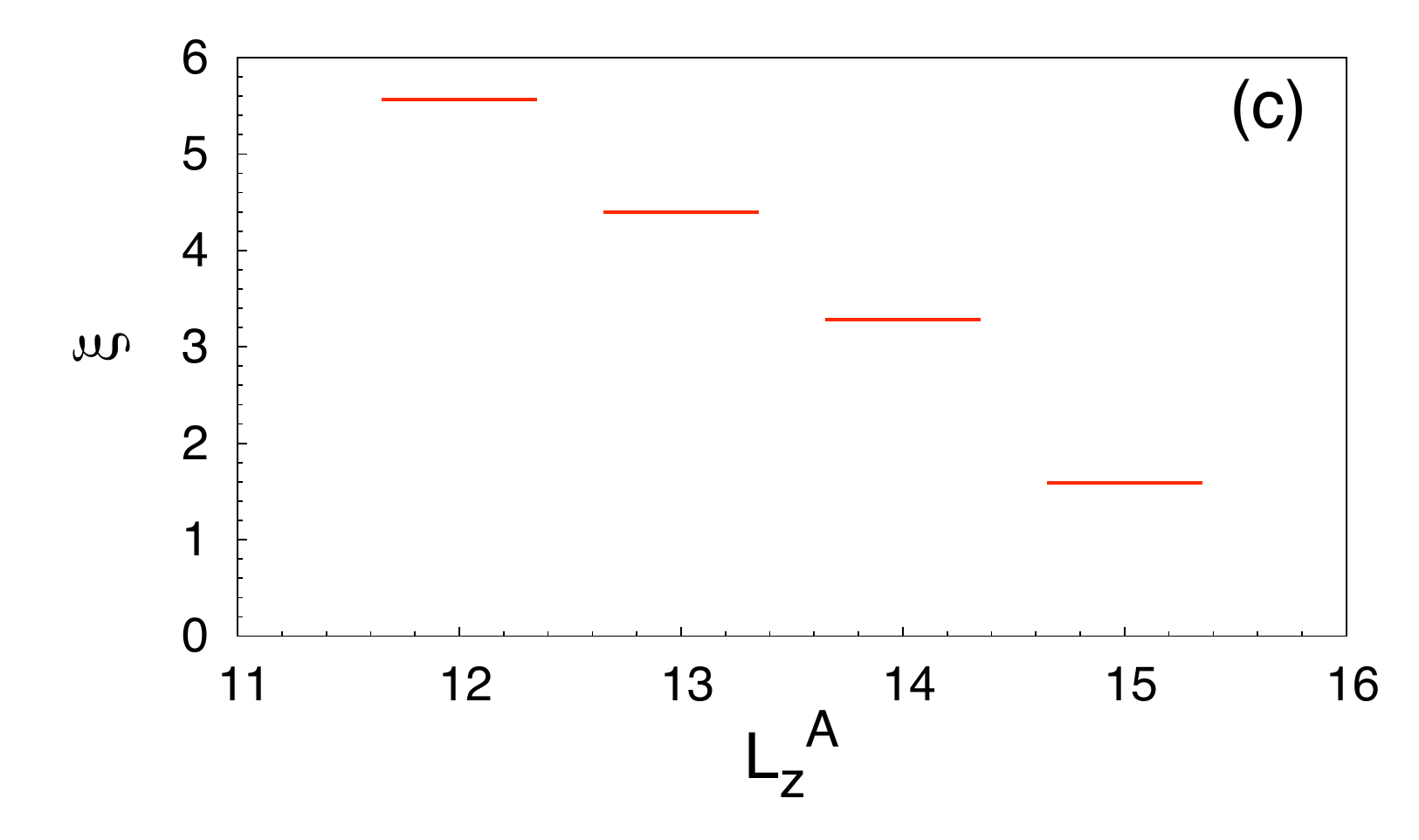}
 \end{minipage}
%\hspace{5pt}
%Laughlin 1/2 big size 12 particles conformal limit
 \begin{minipage}[l]{0.245\linewidth}
   \includegraphics[width=\linewidth]{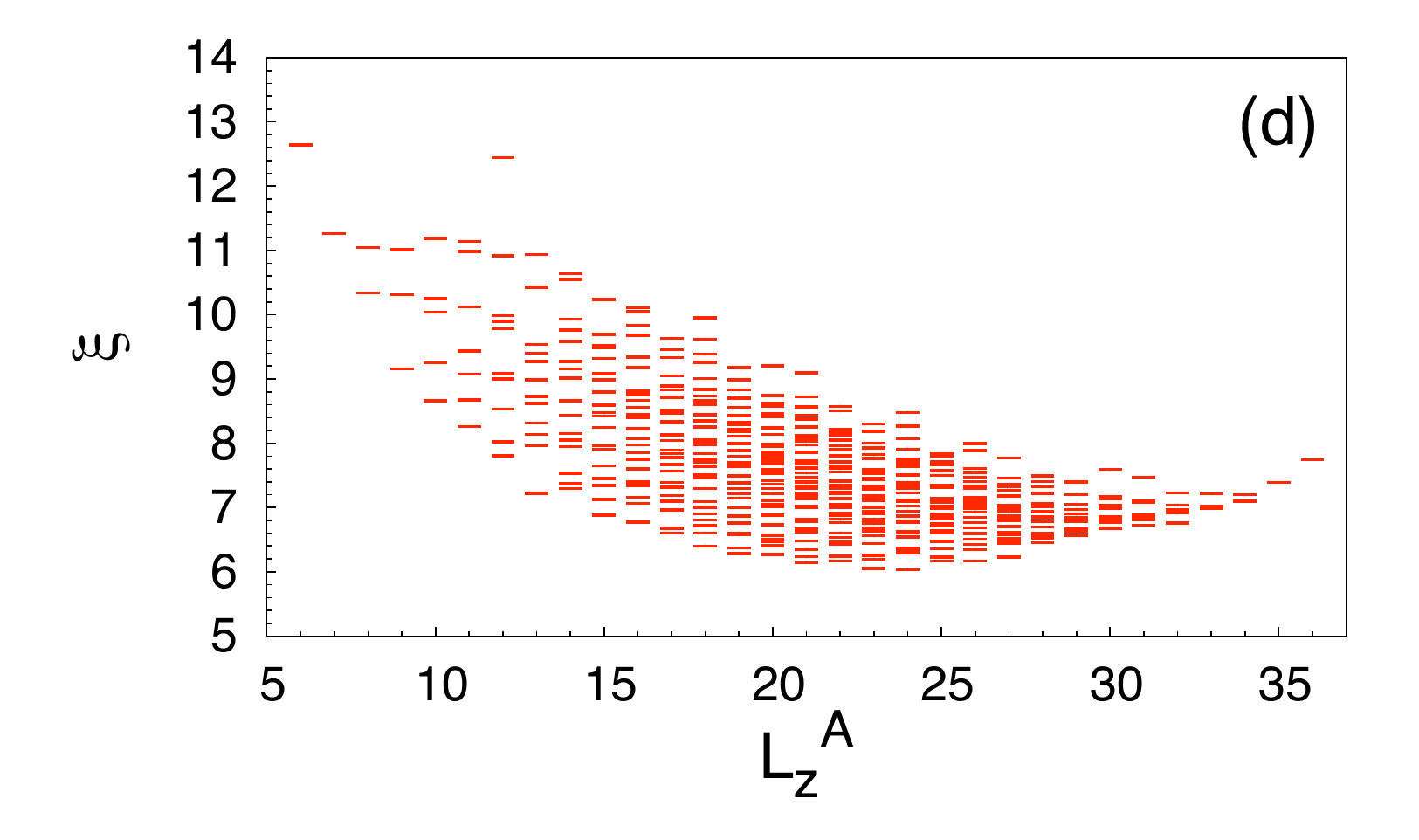}
 \end{minipage}
%\hspace{5pt}
\vspace{-2pt}
\caption{OES of the bosonic $m=2$ Laughlin state for $N=8$ and (a) $N_A=3$, $l_A=7$, (b) $N_A=3$, $l_A=5$, (c) $N_A=3$, $l_A=4$ and OES of the $m=2$ state with $N=12$ and (d) $N_A$=6 and $l_A=11$ in the conformal limit. The counting of plots (a), (b) and (c) should be compared to the number of $(1,2)$-admissible configurations listed in Table~\ref{tab:counting-m2} at each $\Delta L_z$. In all cases, the full counting is predicted exactly by $\mathcal{N}(l_A,N_A, \Delta L_z)$, defined by Eq.~\eqref{eq:genfunc}. }
\label{fig:oem-m2}
\end{figure*}
 %Before stating our conjecture and its properties for general $m$
Let us illustrate Eq.~\eqref{eq:countform}, the counting principle behind the finite-size OES counting of the $1/m$ Laughlin states, with a simple example at $N=8$, $m=2$. 
%The root partition is $n(\lambda_0)=\{101 010101010101\}$.  
Consider the sectors of the OES at $l_A=5$, $N_A=3$. 
As $l_A\leq l_B$, condition (\textbf{i}) is satisfied and we need only consider subsystem $A$. 
We fix $l_{enl,A}$ so that the single occupation number configuration at $L_{z,min}^A$, $\{00003\, |\, \ldots \}$, is identified with exactly one $(1,2)$-admissible configuration at the same angular momentum:
\begin{align}
000\overleftarrow{0}3 |\overrightarrow {\color{red}0}{\color{red}0} \leftrightarrow 00\overleftarrow{0}11 {\color{red}1}|\overrightarrow{\color{red}0}   \leftrightarrow 00101  {\color{red}01}|.
\end{align}
where the hard wall is indicated by `$|$'. 
The arrows indicate angular-momentum conserving operations that result in a (1,2)-admissible configuration. 
Pushing the hard wall further to the right allows for $(1,2)$-admissible configurations at angular momenta lower than $L_{z,min}^A$. 
Thus, $l_{enl,A}=l_A+2=7$. 
 The conjectured counting of the OES is therefore $\mathcal{N}_s(l_A,N_A,\Delta L_z)=\mathcal{N}(5,3,\Delta L_z)$. 
 The middle column of Table~\ref{tab:counting-m2} lists the possible $(1,2)$-admissible configurations of 3 particles in $l_{enl,A}=7$ orbitals at every $\Delta L_z$. 
 The resulting counting, $(1,1,2,2,2,1,1)$, is identical to the counting of the numerically generated OES, Fig.~\ref{fig:oem-m2}(b). 
 We also list some of the $(1,2)$-admissible partitions of $N_A=3$ particles for orbital cuts $l_A=7,4$ ($l_{enl,A}=9,6$), in the first and third column respectively. Their number is identical to the numerically observed level counting shown in Fig.~\ref{fig:oem-m2}(a) and (c).

\begin{table}[htbp]\footnotesize
  \begin{tabular}{l l | l l | l l}
 $l_A=7$, &$N_A=3$  & $l_A=5$, &$N_A=3$& $l_A=4$, & $N_A=3$\\ \hline
$\Delta L_z=0$: & $101010000$  & $\Delta L_z=0$: & $1010100$ & $\Delta L_z=0$: & $101010$ \\
$\Delta L_z=1$: & $101001000$ &  $\Delta L_z=1$: & $1010010$ & $\Delta L_z=1$: & $101001$ \\
$\Delta L_z=2$: & $101000100$ & $\Delta L_z=2$: & $1010001$ & $\Delta L_z=2$: & $100101$ \\
& $100101000$ & 			& $1001010$ & $\Delta L_z=3$: &  $010101$ \\
$\Delta L_z=3$: & $101000010$ & $\Delta L_z=3$: &  $1001001$ & & \\
& $100100100$ & 			& $0101010$ & & \\
& $010101000$ & 		  $\Delta L_z=4$:  & $1000101$ & & \\
$\Delta L_z=4$: & $101000001$ & 			& $0101001$ & & \\
& $100100010$ & 		$\Delta L_z=5$:  & $0100101$	& & \\
& $100010100$ & 		$\Delta L_z=6$:  & $0010101$	& & \\
& $010100100$ & 					& 		& 		& \\
& $\ldots$ & 					& 			& 		&
 \end{tabular}
 \caption{Examples of the finite-size counting of the $m=2$ Laughlin state. The number of $(1,2)$-admissible partitions in $l_A+2$ orbitals at each $\Delta L_z$ equals the number of levels in the OES for the cut with $N_A$ particles in $l_A$ orbitals in Fig.~\ref{fig:oem-m2}.}
  \label{tab:counting-m2}
\end{table}

%\paragraph{General remarks about the rank of the model state}
A few general remarks about the OES counting of the model FQH states are in order. For  a generic state, one expects the rank of the OEM to be equal to the smaller of its dimensions. 
What makes the model FQH states special is the factorially many linear dependencies in their OEMs that keep the rank finite even in the thermodynamic limit. 
The finite-size counting conjectured in this article is expected to be hard to prove in general, although the level counting at both high and low angular momenta $L_z^A$ can be analytically determined. 
The generic U(1) level counting in the OES of $N_A$ particles at  angular momenta $\Delta L_z=0,1,\ldots, N_A^h$ has been rigorously shown for certain orbital cuts \cite{Chandran2011}. 
At $L^A_{z,min}$, $L^A_{z,min} +1, \ldots,L^A_{z,min}+ N_A^h$, the rank of the OEM is the Hilbert space dimension of part $A$. For instance, in the $m=2$ Laughlin state with $N_A=3, l_A=5$, there is only one level in the OES at $L^A_{z,min}$ and $L^A_{z,min}+1$ corresponding to the configurations $\{00003 \,|\,  \ldots \} $ and $\{00012 \,| \,\ldots \}$ of $A$. 
For values of $ L_z^A> L^A_{z,min}+2$, the OEM rank is less than the Hilbert space dimension of $A$, indicative of the nontrivial structure of the OEM.

% FIGURE CHANGED TO N=8 INSTEAD OF N=10

\begin{figure}
 \begin{minipage}[l]{0.49\linewidth}
   \includegraphics[width=\linewidth]{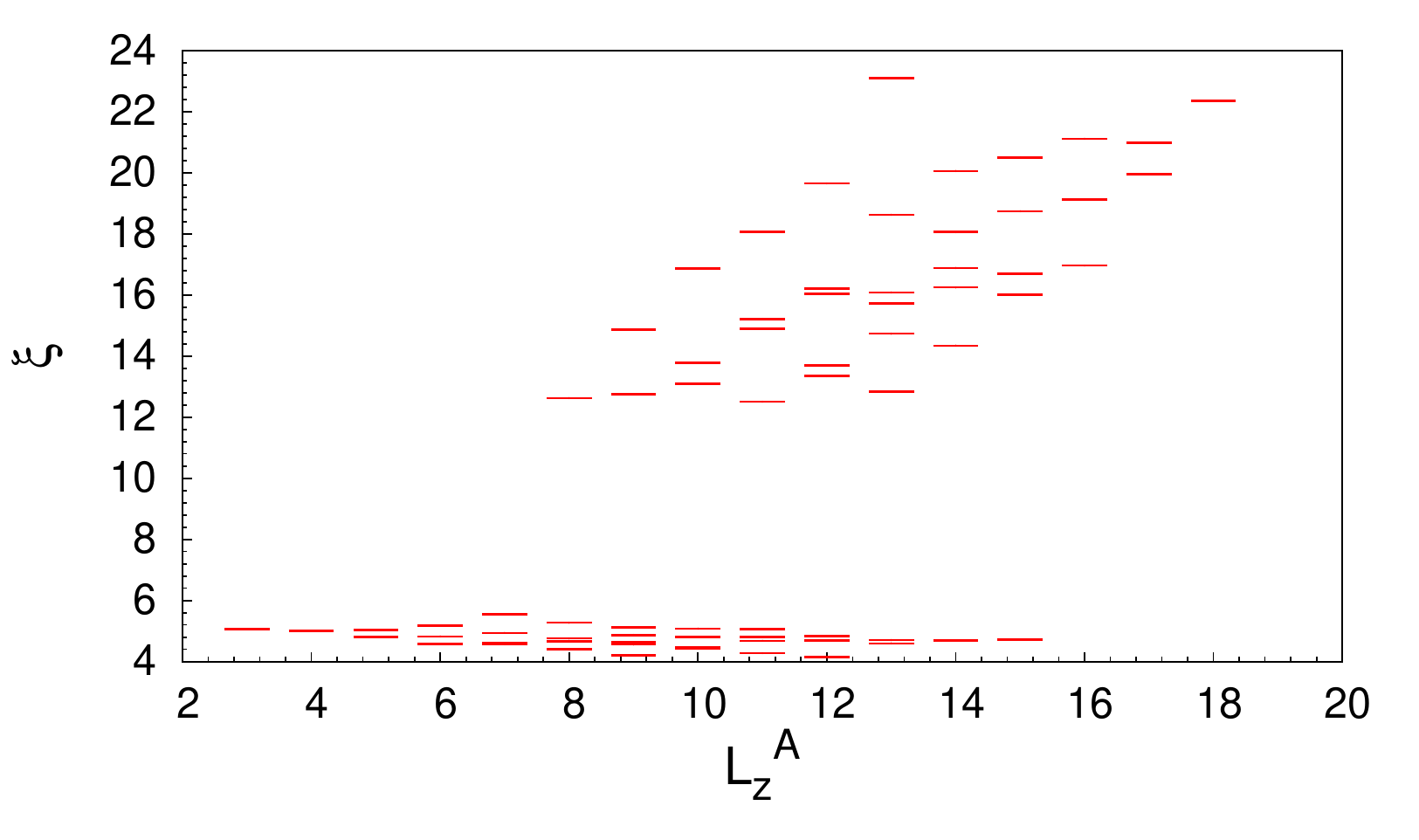}
   %Coulomb 1/2 N=10 N_A=3 l_A=7 full spectrum 
 \end{minipage}
%\hspace{5pt}
 \begin{minipage}[l]{0.49 \linewidth}
   \includegraphics[width=\linewidth]{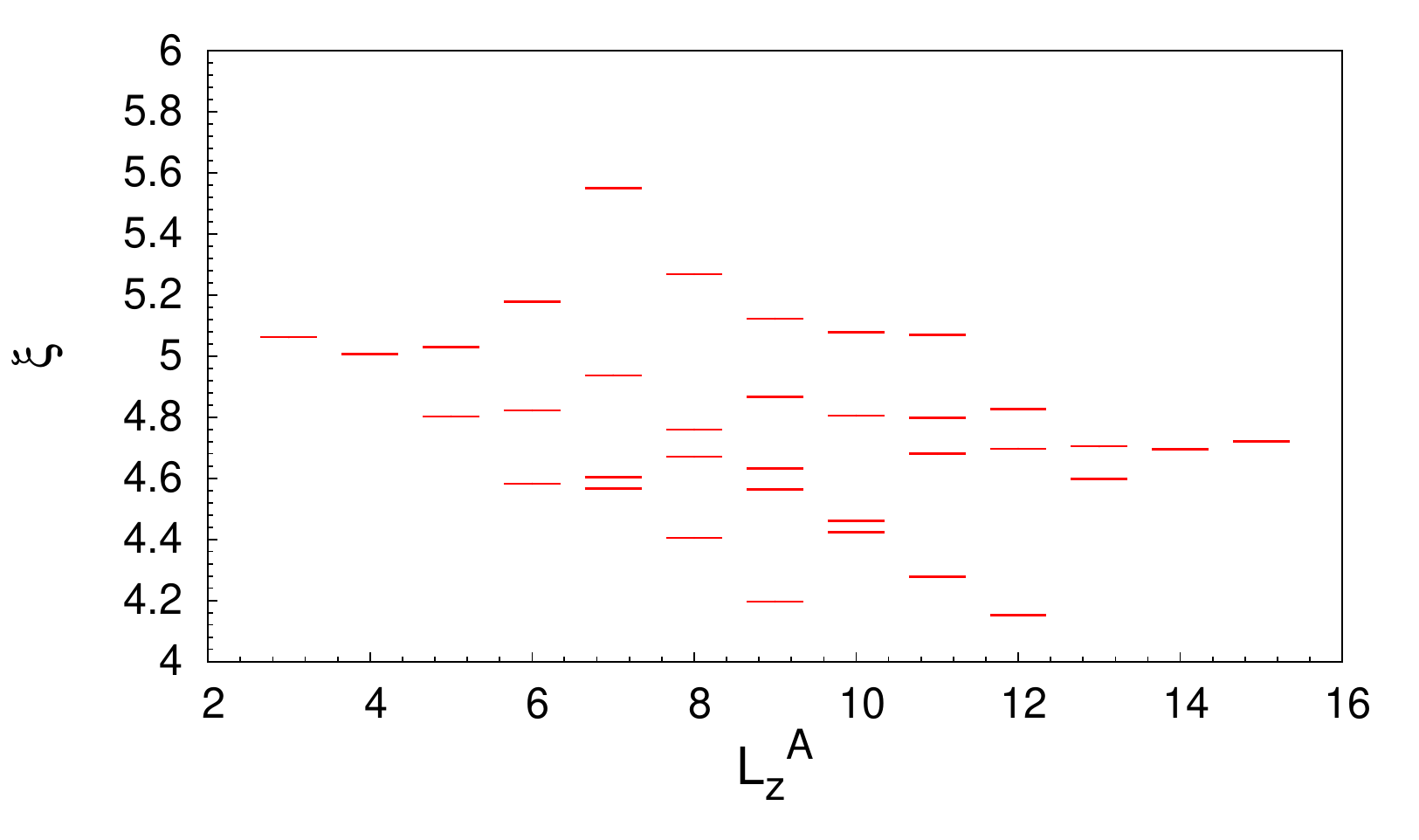}
   %Coulomb 1/2 N=10 N_A=3 l_A=7 low-energy part of spectrum 
 \end{minipage}
 \caption{Coulomb entanglement spectrum of $N=8$ bosons at filling $1/2$ in the sector $l_A=7$, $N_A=3$ in the conformal limit. Left: The full entanglement spectrum with a clear entanglement gap. Right: The low-`energy' part of the conformal entanglement spectrum. The level counting is identical to the Laughlin entanglement spectrum in Fig.~\ref{fig:oem-m2}(a). }
 \label{Fig:Coulombentanglement}
\end{figure}

%\paragraph{Connection to realistic Coulomb ground state}
It is worthwhile to demonstrate the role of our conjecture for the Laughlin states in the OES counting of Coulomb states at the same filling. Fig.~\ref{Fig:Coulombentanglement} shows the entanglement spectrum for the bosonic Coulomb state of $N=8$ particles at filling fraction  $\nu=1/2$ in the sector $l_A=7$, $N_A=3$ in the conformal limit. The low-`energy' part--- shown in the right figure--- is separated from the levels higher in the spectrum by a gap at all $L_z^A$. 
In contrast to the entanglement entropy calculations that rely on scaling arguments, we can determine all the quantum numbers of the edge theory of the state using the conjectured counting of the model OES at a \emph{single} system size and a \emph{single} orbital cut. For instance, the level counting at high $L_z^A$ in Fig.~\ref{Fig:Coulombentanglement} is $(1,1,2,3,5)$ for $\Delta L_z=0,\ldots,  4$, identical to the counting of $3$ particles in any Laughlin state, $\mathcal{N}(l_A,3,\Delta L_z)$, as $l_A\rightarrow \infty$. From Eq.~\eqref{eq:genfunc}, we observe that the \emph{first} discrepancy from the counting at $l_A\rightarrow \infty$ is at $\Delta L_z = N_A^h+1$. In Fig.~\ref{Fig:Coulombentanglement}, this occurs at $\Delta L_z=5$, which fixes $N_A^h=4$. 
Inserting this value into the expression for $N_A^h$, we find that $m=2(l_A-1-N_A^h)/(N_A-1) =2$. 
As the $\Delta L_z$ of the first finite-size correction is the only information required, this analysis works equally well for the cases where the conjecture only provides an upper bound to the counting. 

%\paragraph{Conclusions}
In summary, we conjecture that the counting of the finite-size OES of the $\nu=1/m$ Laughlin states exhibits Haldane exclusion statistics of a boson with compactification radius $\sqrt{m}$ in a box of known orbital length. Our claim is supported by extensive evidence from all numerically accessible sizes. Our conjecture shows that the OES determines all the quantum numbers of the Laughlin state edge theory--- its central charge through the thermodynamic limit counting, and the compactification radius through the finite-size counting. It suggests that the entanglement gap in the Coulomb spectrum protects the Haldane statistics of the phase in the thermodynamic limit and thus provides us with a new way of extracting the boson compactification radius for any state with a finite gap. A natural direction for future research is understanding the counting of the non-abelian states, which is complicated by more quantum numbers. It would also be interesting to extend the analysis in this article to two orbital cuts \cite{calabresecardy2009} on the torus \cite{torus,NAtorus} or the sphere. The resulting OES is expected to be the combination of the finite-size spectra of the two edges if they are non-interacting.

\paragraph{Ackknowledgements}
MH is supported by the Alexander-von-Humboldt foundation, the Royal Swedish Academy of Science and NSF DMR  grant 0952428. BAB is supported by the Sloan Foundation, Princeton startup funds, NSF DMR  grant 0952428, and NSF CAREER DMR-0952428. 
AC, BAB and MH wish to thank Ecole Normale Superieure and Microsoft Station Q for generous hospitality.
BAB thanks the IOP Center for International Collaborations, Beijing, for generous hosting.

\bibliography{finite_oem}

\end{document}